\documentclass[useAMS,usenatbib,usegraphicx]{mn2e}
\usepackage{epsfig}

\title[]{H$_2$O maser emission associated with the planetary nebula IRAS~16333$-$4807}
\author[]{ L. Uscanga,$^{1}$\thanks{E-mail:lucero@astro.noa.gr} J.\,F. G\'omez,$^{2}$ L.\,F. Miranda,$^{2,3}$ P. Boumis,$^{1}$ O. Su\'arez,$^{4}$
\newauthor
J.\,M. Torrelles,$^{5}$ G. Anglada,$^{2}$ and D. Tafoya$^{6}$\\
$^{1}$Institute of Astronomy, Astrophysics, Space Applications and Remote Sensing, National Observatory of Athens, \\
15236 Athens, Greece\\
$^{2}$Instituto de Astrof\'isica de Andaluc\'ia, CSIC, Apartado 3004, 18080 Granada, Spain\\
$^{3}$Departamento de Fisica Aplicada, Facultade de Ciencias, Universidade de Vigo, 36310 Vigo, Spain\\
$^{4}$Laboratoire Lagrange, UMR7293, Universit\'e de Nice Sophia-Antipolis, CNRS, Observatoire de la C\^ote d'Azur, \\
06304 Nice Cedex 4, France \\
$^{5}$Institut de Ci\`encies de l'Espai (CSIC-IEEC) and Institut de Ci\`encies del Cosmos (UB-IEEC), Mart\'i i Franqu\`es 1, \\
08028 Barcelona, Spain\\
$^{6}$Centro de Radioastronom\'ia y Astrof\'isica, UNAM, Apartado Postal 3-72 (Xangari), 58089 Morelia, Mexico}

\begin{document}


\pagerange{\pageref{firstpage}--\pageref{lastpage}} \pubyear{2014}

\maketitle

\label{firstpage}

\begin{abstract}
We present simultaneous observations of H$_2$O maser emission and radio continuum at 1.3\,cm carried out with the Australia Telescope Compact Array towards two sources, IRAS 16333$-$4807 and IRAS 12405$-$6219, catalogued as planetary nebula (PN) candidates, and where single-dish detections of H$_2$O masers have been previously reported. Our goal was to unambiguously confirm the spatial association of the H$_2$O masers with these two PN candidates. 
We detected and mapped H$_2$O maser emission in both fields, but only in IRAS 16333$-$4807 the maser emission is spatially associated with the radio continuum emission.
The properties of IRAS 16333$-$4807 provide strong support for the PN nature of the object, hereby confirming it
as the fifth known case of a H$_2$O maser-emitting PN.
This source is bipolar, like the other four known H$_2$O maser-emitting PNe, 
indicating that these sources might pertain to a usual, but short phase in the evolution of bipolar PNe.
In IRAS 12405$-$6219, the H$_2$O maser and radio continuum emission are not associated with each other and, in addition, the available data indicate that this source is an H\,{\sc ii} region rather than a PN.
\end{abstract}


\section{Introduction}

Planetary nebulae (PNe) form at the end of the evolution of low- and intermediate- mass stars ($\le 8$~M$_{\sun}$), after traversing the asymptotic giant branch (AGB) phase and a brief ($\simeq$1000 yr) post-AGB phase. 
During these last phases of evolution, maser emission of different molecules (e.g., H$_2$O, OH, SiO) is found in the envelope of oxygen-rich stars \citep{ric12}. The presence of maser emission is indicative of energetic processes, such as strong radiation fields and shocks. In the particular case of H$_2$O masers, this emission is common in
 oxygen-rich AGB stars and originates in the inner regions of the envelope, between 10 and 100 au from the central star \citep{rei81}. 

It was initially thought that H$_2$O masers disappear within $\sim$100 yr after the cessation of the strong mass loss ($\sim$10$^{-5}$ M$_{\sun}$~yr$^{-1}$) typical of the AGB phase \citep{lew89,gom90}. This timescale implies that it is possible to find H$_2$O masers during the post-AGB phase \citep{eng02}, albeit with a lower detection rate than in the AGB phase, but they were not expected to be found after entering the PN phase.
However, four PNe have been found so far to harbour H$_2$O masers: IRAS 19255$+$2123 (K3$-$35) \citep{mir01}, IRAS 17347$-$3139 \citep{degreg04}, IRAS 18061$-$2505 \citep{gom08}, and IRAS 15103$-$5754 \citep{sua12}.

All these H$_2$O maser-emitting PNe (hereafter H$_2$O-PNe) exhibit a bipolar morphology in the optical, infrared and/or radio continuum images
with signs of point-symmetry, suggesting that precessing bipolar jets could be involved in their formation \citep{vel07,lag11}.
In these objects, the H$_2$O masers are usually located towards the central region (within $\sim$200 au), mostly aligned in a direction perpendicular to the main nebular axis defined by the bipolar lobes \citep{mir10}. 
This strongly suggests that these masers are related to dense equatorial structures (discs or toroids) around the central star \citep{usc08},
although in IRAS 19255$+$2123 (K3$-$35) H$_2$O maser emission was also detected at a few thousand au from the centre associated with the tips of the precessing bipolar jets in the object \citep{mir01}. 
In IRAS 15103$-$5754 the maser emission shows a completely different pattern, since it mainly traces a high-velocity ($\sim$80 km~s$^{-1}$), collimated jet \citep{sua12}.
This pattern allows the classification of IRAS 15103$-$5754 as a ``water fountain'', a class of sources that 
mainly includes post-AGB stars \citep{ima07}.
Given the different structures traced by the maser emission,
PNe with H$_2$O masers are key to understand the change in the mass-loss regime in the transition from the AGB to the PN phase.
Furthermore, 
H$_2$O-PNe are most probably among the youngest PNe and, therefore, their study and the confirmation of new cases are crucial to understand the earliest stages of PN evolution.

\citet{sua07} and \citet{sua09} performed a single-dish survey of H$_2$O masers towards post-AGB stars and young PN candidates. Their goal was to relate the presence of H$_2$O masers with the evolutionary stage and the characteristics of the sources observed. 
The observations of the northernmost sources (Su\'arez et al. 2007) led to the detection of H$_2$O maser emission towards the PN IRAS 18061$-$2505, later confirmed with the Very Large Array \citep{gom08}. For the southern part of the survey \citep{sua09}, two additional sources catalogued as possible PNe, IRAS 16333$-$4807 and IRAS 12405$-$6219 \citep{van93}, were found to be probably associated with H$_2$O maser emission.
However, the lack of enough angular resolution of the single-dish observations prevented to establish a clear association.

In this paper we present simultaneous observations of H$_2$O maser emission and radio continuum at 1.3 cm 
carried out with the Australia Telescope Compact Array (ATCA)\footnote{The ATCA is part of the Australia Telescope, which is funded by the Commonwealth of Australia for operation as a National Facility managed by CSIRO}
towards IRAS 16333$-$4807 and IRAS 12405$-$6219. Our main goal was to confirm the association of H$_2$O maser emission, previously detected with the Parkes antenna \citep{sua09}, with these two PN candidates. 

\section[]{Observations}

Observations were carried out with the ATCA in its 6A configuration on 2013 February 12. The Compact Array Broadband Backend (CABB) of ATCA provides two independent IF bands of $\simeq 2$ GHz each, in dual linear polarisation. In our case, these two bands were centred at 22 and 24 GHz. We used the 64M mode of CABB, which samples each IF band into 32 wideband channels of 64 MHz each. The wideband channel containing the $6_{16}-5_{23}$ transition of the water molecule (rest frequency = 22235.08 MHz) was then zoomed in, by sampling it into 2048 channels, which provides a velocity resolution of $\simeq$0.42 km s$^{-1}$, and a total coverage of  $\simeq$860 km s$^{-1}$ for the spectral line. 

A summary of the observations is given in Table 1.
We used PKS 1253-055 as the bandpass calibrator. 
The absolute flux density scale was set using PKS 1934-638 as the reference. 
The uncertainty in the flux calibration is $\simeq10$\%.
All broadband (continuum) data were calibrated and imaged with the MIRIAD package. To obtain the continuum maps presented in the figures of this paper, we processed together the whole 4 GHz frequency coverage, flagging out the broadband channel containing the maser emission.
Dirty maps of the continuum emission were obtained with task ``invert'' of MIRIAD, applying frequency synthesis and a robust parameter of 0.5.
Images were subsequently deconvolved with the CLEAN algorithm, as implemented in task ``mfclean''. Self-calibration was applied to these continuum images to improve their dynamic range. 
Final rms noise in the continuum maps were $\simeq 40$ $\mu$Jy~beam$^{-1}$.

\begin{table*}
\begin{minipage}{170mm}
\caption{Summary of Observations}
\begin{tabular}{@{}ccccccc}
\hline
Source & Coordinates & Phase & $\theta_c$\footnote{Half-power width of the synthesized beam in the broad-band continuum maps} & P.A.\footnote{Position angle (north to east) of the major axis of the synthesized beam in the broad-band continuum maps} & $\theta_l$\footnote{Half-power width of the synthesized beam in the spectral line maps} &  P.A.\footnote{Position angle of the major axis of the synthesized beam in the spectral line maps}\\
& (J2000) & Calibrator & (arcsec) & (deg) & (arcsec) & (deg)\\
\hline
IRAS 16333$-$4807 & 16$^{\mathrm{h}}$37$^{\mathrm{m}}$06.63$^{\mathrm{s}}$, $-$48$^{\mathrm{o}}$13$'$42.5$''$ & PMN J1650-5044 & 0.60$\times$0.38 & 11 & 0.90$\times$0.34   & 27 \\
IRAS 12405$-$6219 & 12$^{\mathrm{h}}$43$^{\mathrm{m}}$31.50$^{\mathrm{s}}$, $-$62$^{\mathrm{o}}$36$'$14.0$''$ & PKS J1147-6753 & 0.53$\times$0.36 & 3 & 0.66$\times$0.34 & $-$6\\ 
\hline
\end{tabular}
\end{minipage}
\end{table*}

In the case of the spectral line data, initial calibration and imaging was also carried out with MIRIAD. Then, we selected the spectral line channel with the highest flux density, and applied self-calibration with the AIPS package of NRAO. Amplitude and phase corrections obtained were then copied to the whole line data set. A new continuum data set at $\simeq 22$ GHz was obtained by averaging the line-free channels. Although the sensitivity of this continuum data was lower than that in the broadband CABB data, it has the advantage of sharing the same corrections obtained from the self-calibration of the maser line. Therefore, the systematic astrometric errors and those due to atmospheric phase variations are the same in line and continuum. 
Deconvolved maps of line and narrow-band continuum emission were obtained with task IMAGR of AIPS, applying a robust parameter of 0.
The typical rms for individual spectral line channels is $\simeq1800$ $\mu$Jy beam$^{-1}$ while the rms for the narrow-band continuum is $\simeq59$ $\mu$Jy beam$^{-1}$.
In this way, we can ascertain the relative position between the maser components and the maximum of the continuum emission with a very high accuracy, which is only limited by the noise in the images, in the form $\sigma_{\rm pos}=\theta/(2\times snr)$, where $\sigma_{\rm pos}$ is the $1\sigma$ relative positional accuracy, $\theta$ is the size of the synthesized beam, and $snr$ is the signal-to-noise ratio of the emission. In our case, the relative accuracies between the maser and the continuum emission were $\simeq 0.4$ and $0.8$ mas for IRAS 16333$-$4807 and IRAS 12405$-$6219, respectively. Continuum emission was subtracted from the line data with task UVLIN of AIPS.

\section[]{Individual sources and observational results}

\subsection{IRAS 16333$-$4807}
Radio continuum and H$_2$O maser emission have been detected towards IRAS \mbox{16333$-$4807}. Fig. 1 presents the H$_2$O maser spectrum while Fig. 2 shows the radio continuum map at 1.3\,cm superimposed to the mid-IR image in the [Ne\,{\sc ii}] filter (12.8\,$\mu$m) by \citet{lag11} as well as the location of the H$_2$O maser components.

The radio continuum map at 1.3\,cm shows a bright core with two lobes that extend $\simeq$2.4 arcsec from the centre. The peak of the radio continuum emission is located at RA(J2000)$=16^{\mathrm{h}}$$37^{\mathrm{m}}$$06.601^{\mathrm{s}}$, Dec(J2000)$=-48^{\mathrm{o}}13'42.88''$ ($\pm0.09''$) with a flux density of 97$\pm$10 mJy. This morphology is very similar to that observed in the [Ne\,{\sc ii}] filter, which shows a bipolar object with two prominent lobes and an unresolved central core defining a very narrow waist. The bipolar lobes present a slight point-symmetry (see \citealt{lag11}).

The H$_2$O maser emission (Fig. 1) presents two blended velocity components, with flux densities of 4.8$\pm$0.5 Jy ($V_{\mathrm{LSR}}\simeq-$41.8 km~s$^{-1}$) and 1.26$\pm$0.13 Jy ($V_{\mathrm{LSR}}\simeq-$43.9 km~s$^{-1}$). 
The stronger component is located at
RA(J2000)=$16^{\mathrm{h}}37^{\mathrm{m}}06.593^{\mathrm{s}}$, Dec(J2000)=  \mbox{$-48^{\mathrm{o}}13'42.82''$} ($\pm0.09''$). The H$_2$O masers are located $\simeq$0.1 arcsec northwest from the peak of the radio continuum (Fig. 2), indicating a spatial association with the object. We note that the systemic velocity of the object is unknown and, therefore, we cannot establish whether the H$_2$O masers arise in an equatorial structure or in a collimated outflow. Finally, the flux density of the H$_2$O maser emission has increased from 1.8 Jy in 2007 (Parkes observations, \citealt{sua09}) to 4.8 Jy in 2013 (ATCA observations, this paper).

\begin{figure}
\includegraphics[width=82mm]{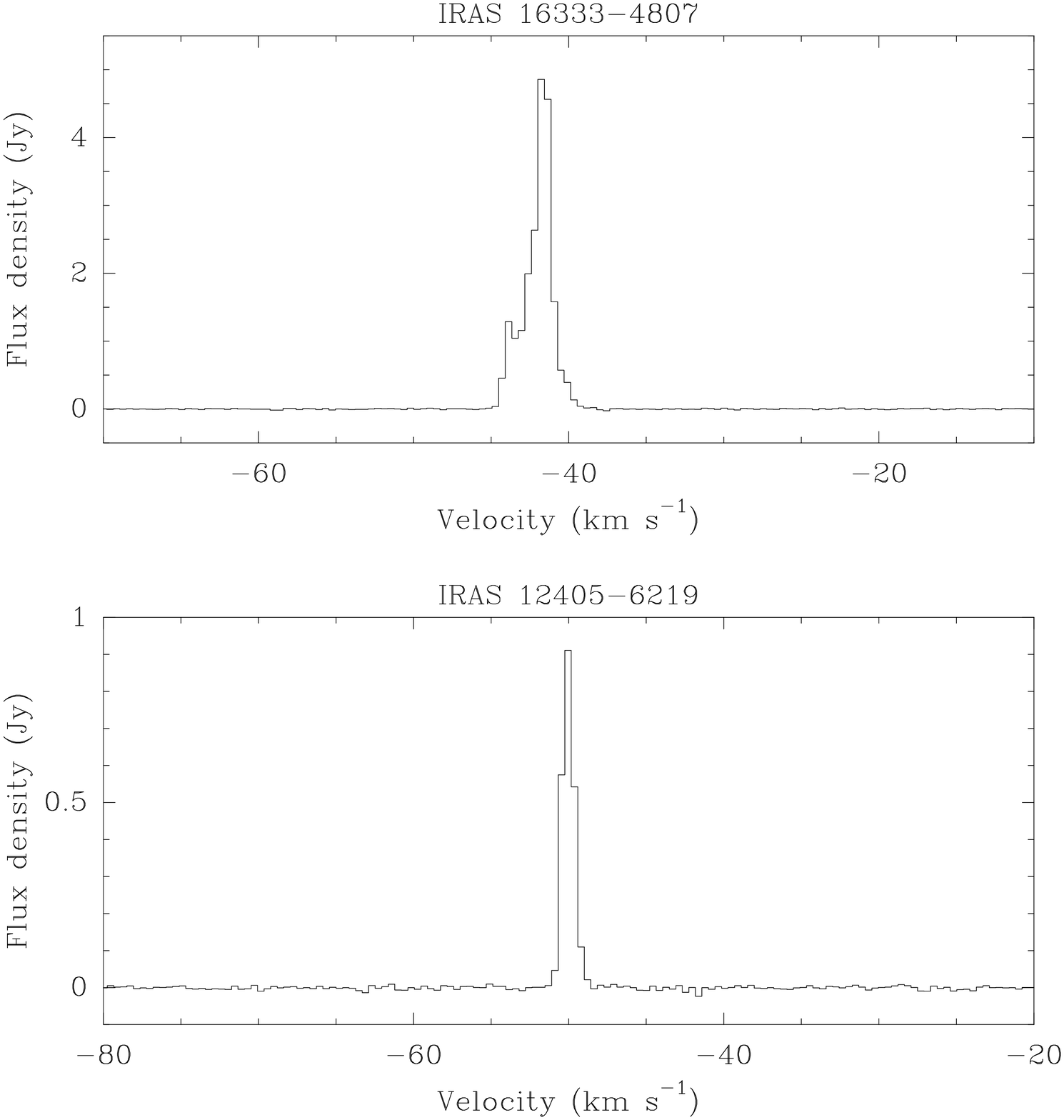}
 \caption{H$_2$O maser spectra towards IRAS 16333$-$4807 and IRAS 12405$-$6219.} 
\end{figure}

\begin{figure*}
\centering
\begin{tabular}{cc}
\epsfig{file=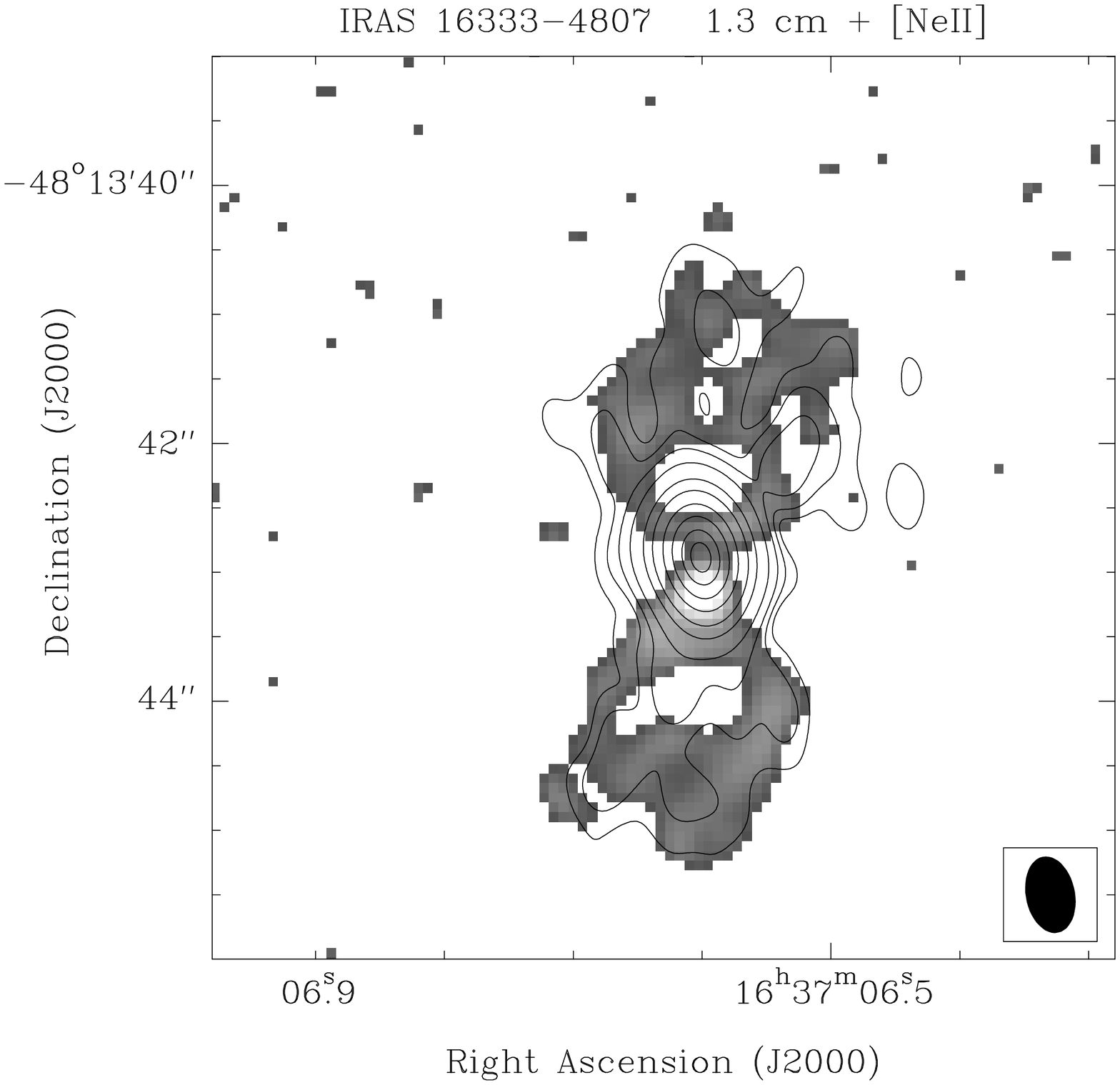,width=77mm,clip=true} & 
\epsfig{file=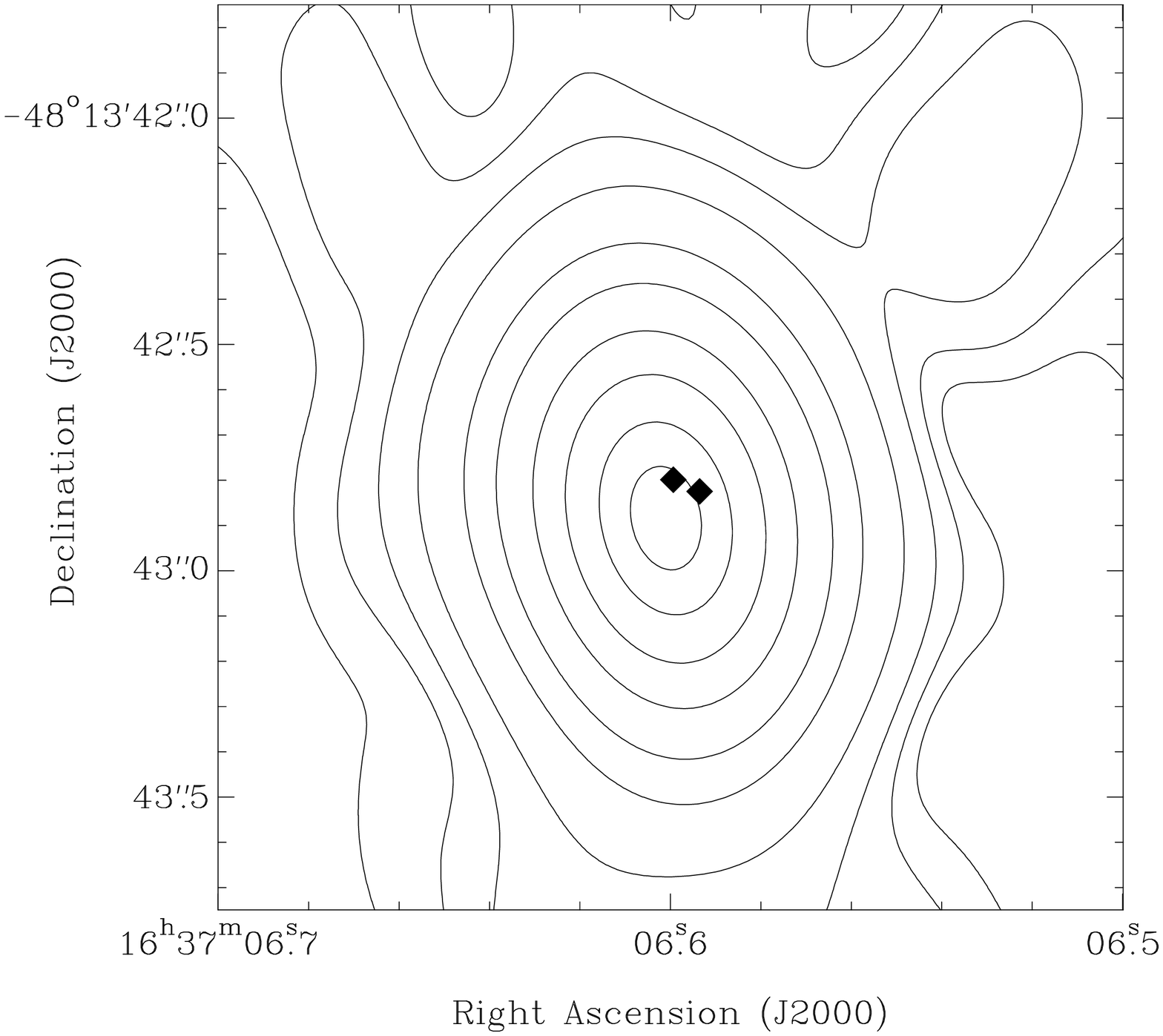,width=80mm,clip=true} 
\end{tabular}
 \caption{\textit{Left:} Contour map of the radio continuum emission of IRAS 16333$-$4807 at 1.3\,cm from ATCA (this paper), superimposed to a mid-IR image in the [Ne\,{\sc ii}] filter (12.8$\mu$m) \citep{lag11}. The contours are 5, 10, 20, 50, 150, 300, 600, 1000, 1500, and 1900 times $3.6\times10^{-5}$ Jy beam$^{-1}$.
The synthesized beam of the radio continuum image is shown at the bottom right corner as a filled ellipse.
\textit{Right:} Close-up of the central nebular region showing the radio continuum emission (contours, same as in the left panel) and the location of the H$_2$O maser components (filled diamonds). The maser component with the maximum flux density is the westernmost one. 
}
\end{figure*}

\subsection{IRAS 12405$-$6219}
We detected H$_2$O maser emission with a single spectral component (Fig. 1) at $V_{\mathrm{LSR}}\simeq-$50 km~s$^{-1}$ and a flux density of 
910$\pm$90 mJy located at RA(J2000)$=12^{\mathrm{h}}43^{\mathrm{m}}32.319^{\mathrm{s}}$, Dec(J2000)$=-62^{\mathrm{o}}$ $36'$ $16.80''$ ($\pm0.07''$). 
A source of radio continuum emission at 1.3\,cm is detected with a peak position at RA(J2000)$=12^{\mathrm{h}}$$43^{\mathrm{m}}$$31.496^{\mathrm{s}}$, Dec(J2000)$=-62^{\mathrm{o}}36'13.40''$ ($\pm0.07''$) and flux density of 
102$\pm$10 mJy. 
This radio continuum emission seems to be associated with the near-IR emission, which is located at RA(J2000)$=12^{\mathrm{h}}$$43^{\mathrm{m}}$$31.52^{\mathrm{s}}$, Dec(J2000)$=-62^{\mathrm{o}}36'13.55''$ (source 2MASS J12433151-6236135 in the 2MASS Point Source Catalog, $2\sigma$ positional uncertainty $\simeq 0.18''$). 
The radio continuum emission shows a cometary morphology (Fig. 3) similar to that of the infrared emission \citep{lag11}.
The H$_2$O maser is located $\sim$6 arcsec southeast from the radio continuum source, showing that these emissions are not spatially associated.

Images from the Wide-field Infrared Survey Explorer (WISE) database show mid-IR emission that is dominated at all bands by the radio continuum source, but at the longest WISE wavelength (22 $\mu$m) there is a weak extension towards the location of the H$_2$O maser. This suggests that the maser is probably associated with a different, highly obscured source.

As already mentioned, IRAS 12405$-$6219 was classified as a PN candidate from its radio continuum emission \citep{van93}. However, \citet{lag11} and \citet{ram12} argue that the object is an H\,{\sc ii} region. 
In particular,
the cometary morphology observed at 1.3\,cm and mid-IR images is 
unusual in PNe but common to H\,{\sc ii} regions.
The object is associated with extended and filamentary emission in near-IR images,
as typically observed in young stellar objects (YSOs).
These properties strongly support the classification of IRAS 12405$-$6219
as an H\,{\sc ii} region.
Regardless of the exact nature of the source, we can certainly rule out that the object is an H$_2$O-PN, given the non-association between the maser and radio continuum emission.

\begin{figure}
\includegraphics[width=82mm]{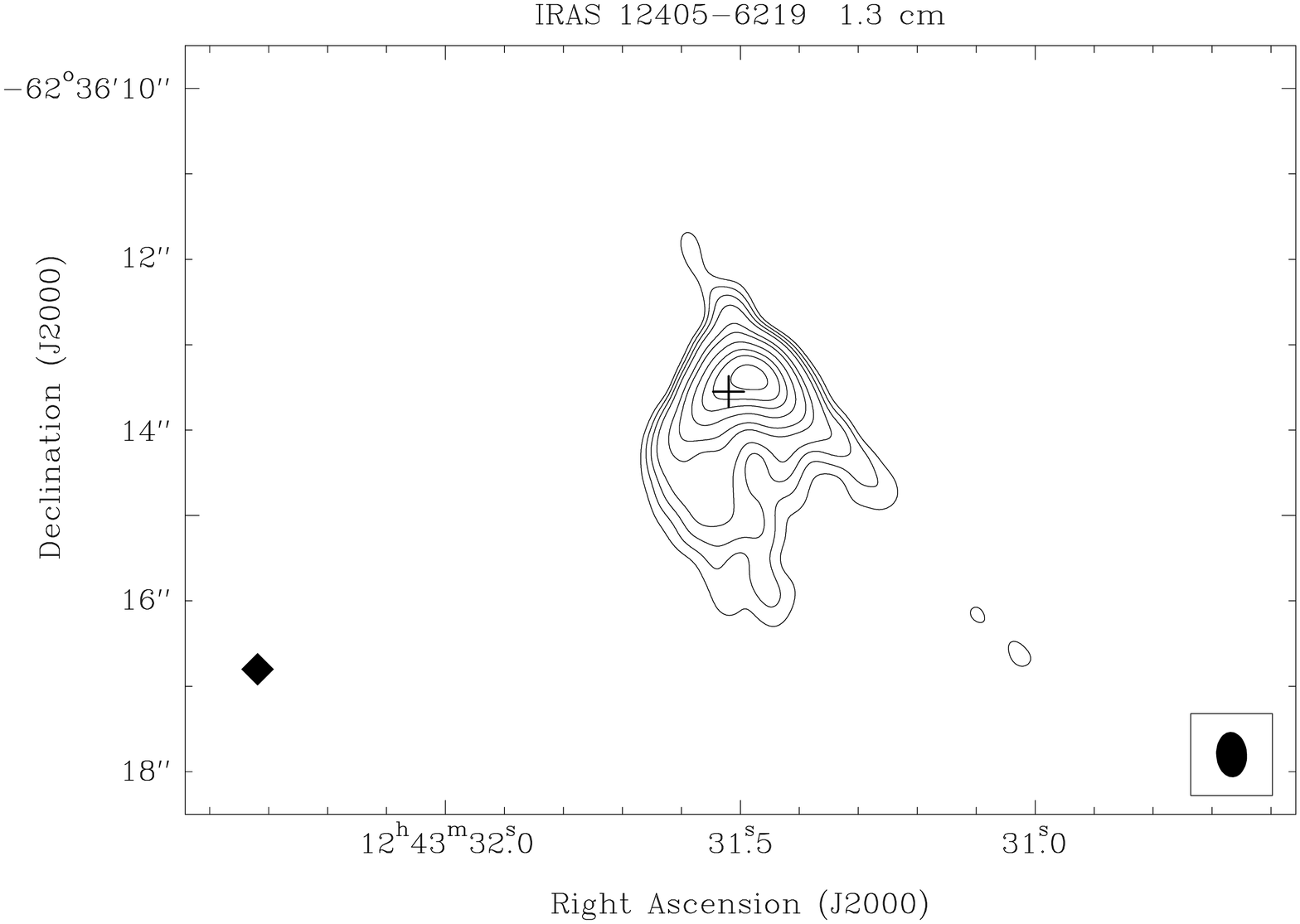}
 \caption{Contour map of the radio continuum emission of IRAS 12405$-$6219 at 1.3\,cm from ATCA. 
The contours are 8, 15, 25, 40, 80, 120, 200, 300, 400, and 550 times $4.0\times10^{-5}$ Jy beam$^{-1}$.
The cross indicates the position of the infrared source 2MASS J12433151-6236135. The size of the cross represents the $2\sigma$ positional uncertainty.
The position of the H$_2$O maser, which is clearly not associated with the continuum emission, is indicated by a filled diamond located near the bottom left corner of the figure. The synthesized beam is shown at the bottom right corner.
}
\end{figure}

\section{Discussion}
\subsection{The nature of IRAS 16333$-$4807 as a planetary nebula}

The presence of radio continuum emission and the infrared colours of
this source are compatible with it being a PN but it could also be
a massive YSO. 
In fact, it was included in the Red MSX Source (RMS) survey
\citep{lum13}
 as a candidate massive YSO,
based on its MSX colours. The RMS
database\footnote{http://rms.leeds.ac.uk/cgi-bin/public/RMS\_DATABASE.cgi},
however, indicates that it is a PN. 
The source is relatively isolated in infrared and submillimetre
images (see, e.g., \citealt{ram12} and the RMS database). Specially in
the case of mid- and far- infrared images, YSOs are associated
with extended and filamentary structures, which are not present in
IRAS 16333$-$4807. Moreover, as shown in the RMS database, there is no
$^{13}$CO emission towards the source \citep{urq07}, whose presence would 
be expected 
in the case of a YSO embedded in a molecular cloud. This
confirms that the source is an evolved object.

IRAS 16333$-$4807 exhibits a distinct bipolar
morphology in the near- \citep{ram12}, mid-IR \citep{lag11}, 
and radio continuum images (Fig. 2). This morphology is common in
post-AGB stars and PNe, but it is rare in H\,{\sc ii} regions. To ascertain its
nature as a PN, as opposed to being a post-AGB star, we have to confirm
the presence of photoionization. Considering the flux density we
determined at $\simeq 23$ GHz ($\simeq 97$ mJy) and that obtained by
 \cite{van93} at 5 GHz (32.5 mJy), we estimate a spectral index of
 $\simeq 0.71$ for the radio continuum emission between those
 frequencies. This spectral index is similar to the one obtained for
 the H$_2$O-PN IRAS 17347$-$3139 \citep{gom05} and clearly 
indicates that the radio continuum is of
 thermal origin. Although such spectral indices may also arise from
 ionized jets, the flux density in IRAS 16333$-$4807 is too high to be
 explained in this way with reasonable mass-loss rates. As in the case of IRAS
 17347$-$3139 \citep{degreg04}, a mass loss rate of $\dot{M}\simeq
 10^{-4} (D/{\rm kpc})^{3/2}$ M$_\odot$ yr$^{-1}$, where $D$ is the distance 
to the source,
 would be required for a spherical ionized wind to explain the observed radio
 continuum emission \citep{wri75,pan75}. 
Even assuming a collimated wind, and following the formulation by \citet{rey86}, the required mass loss rate is as high as $\simeq 3\times 10^{-6} (D/{\rm kpc})^{3/2}$ M$_\odot$ yr$^{-1}$, and this is a lower limit, assuming that the wind is fully ionized.
Despite the uncertainty in the distance, such mass loss rates
 are significantly higher than that expected from the central stars of
 PNe and post-AGB stars ($\leq 10^{-7}$ M$_\odot$ yr$^{-1}$; \citealt{pat91}; 
\citealt{vas94}). 
Therefore, the only remaining possibility to explain the observed radio continuum emission in IRAS 16333$-$4807 is that it arises from a photoionized region, confirming that the object is a PN.

\subsection{Comparison with other H$_2$O maser-emitting PNe}

The main result of this paper is the confirmation of IRAS 16333$-$4807
as a H$_2$O-PN. This is the fifth object that is considered a
bona fide H$_2$O-PN. As mentioned above, it was previously thought
that H$_2$O maser emission could not be present in PNe. However, a
population of such sources is now being uncovered. The presence of H$_2$O
masers in PNe may not be such a rare case after all, but it could
represent a usual but short phase in the evolution of some types of PNe.

Although the known number of this type of sources is still scarce,
such number is growing, and now we can start to study them as a group. 
Their most
outstanding common characteristic is that all five H$_2$O-PNe have
bipolar morphologies in their infrared and/or optical
images. So, H$_2$O-PNe might pertain to a usual phase in the evolution of
bipolar PNe.

Other than that, among the
five H$_2$O-PNe there are some interesting differences, although there
is a relatively homogeneous subgroup of three of them (IRAS 19255$+$2123 (K3$-$35), 
IRAS 17347$-$3139, and IRAS
16333$-$4807) that may be considered as representative. These three
sources are all strongly obscured, and their H$_2$O maser emission
appears close to
the central star, probably tracing a circumstellar torus. 
We note that the H$_2$O masers at the tips of the bipolar lobes 
observed in IRAS 19255$+$2123 (K3$-$35)
have not been seen in subsequent observations of the source
\citep{degreg04}
which suggests that this particular characteristic could be a transient 
phenomenon in
H$_2$O-PNe. Moreover, these are the only three H$_2$O-PNe in which the
photoionized gas traced by the radio continuum emission shows a
bipolar morphology similar to that seen in optical and infrared
images. 

\citet{gom08} and \citet{sua09} proposed a simple evolutionary schema for H$_2$O masers in evolved stars (5$-$8 M$_{\sun}$), considering their location in the nebula and their radial velocities. 
First, in the AGB phase, low-velocity H$_2$O maser emission arises
from the spherical shell. Later on, during the late AGB or post-AGB
phase, the anisotropic mass-loss starts, which is characterised by
H$_2$O masers tracing high-velocity jets, marking the entrance in the
``water fountain'' phase. In some water fountains there are also masers tracing
equatorial structures \citep{ima07}, in addition to the ones along jets.
The next phase would be the one in which water fountain jets are quenched, while H$_2$O masers survive only in dense equatorial regions. 

Considering this evolutionary schema, among the five known H$_2$O-PNe,
IRAS 15103$-$5754 \citep{sua12} would be clearly younger than
the rest, since its maser emission still traces the collimated jets
typical of  ``water
fountains''.  The subgroup of H$_2$O-PNe, IRAS   \mbox{19255$+$2123} (K3$-$35),
IRAS 17347$-$3139, and IRAS 16333$-$4807, would be in a somewhat later
evolutionary stage. Finally, IRAS 18061$-$2505 would be more evolved
since it is completely visible in the optical
\citep{sua06}, showing at this wavelength the central star and two bipolar lobes with a total extension of $\simeq 45$ arcsec ($\simeq 59000$ au at the distance of 1.3 kpc).
 
\section{Conclusions} 
In IRAS 16333$-$4807, the detected H$_2$O masers are associated with the radio continuum emission at 1.3\,cm and are observed towards the centre of the object. The properties of IRAS 16333$-$4807 indicate a PN nature, making it the fifth known PN with H$_2$O maser emission. The object shows a bipolar morphology, as the other four H$_2$O-PNe. This result further supports the idea that H$_2$O masers in PNe are not the remnant of masers in the AGB phase 
but are excited by energetic phenomena associated with the formation of PNe.

In IRAS 12405$-$6219, the H$_2$O maser emission is not associated with the radio continuum emission at 1.3\,cm. The available data strongly indicates that this object is an H\,{\sc ii} region rather than a PN.

\section*{Acknowledgements}
LU acknowledges support from grant PE9-1160 of the Greek General Secretariat for Research and Technology in the framework of the program Support of Postdoctoral Researchers. GA, JFG, LFM, and JMT acknowledge support from MICINN (Spain) AYA2011-30228-C03 grant (co-funded with FEDER funds). The ICC (UB) is a CSIC-Associated Unit through the ICE (CSIC).
LU acknowledges the hospitality and support of the staff at the Paul Wild Observatory in Narrabri (NSW) during the observations.
This paper made use of information from the Red MSX Source survey database at http://rms.leeds.ac.uk/cgi-bin/public/RMS\_DATABASE.cgi which was constructed with support from the Science and Technology Facilities Council of the UK. It also made use of data products from the Wide-field Infrared Survey Explorer, which is a joint project of the University of California, Los Angeles, and the Jet Propulsion Laboratory/California Institute of Technology, funded by the National Aeronautics and Space Administration.

\bsp

\label{lastpage}

\end{document}